\pdfoutput=1
\PassOptionsToPackage{unicode}{hyperref}
\PassOptionsToPackage{hyphens}{url}
\PassOptionsToPackage{dvipsnames,svgnames,x11names}{xcolor}
\documentclass[
  10pt,
]{article}
\usepackage{xcolor}
\usepackage[margin=2.2cm]{geometry}
\usepackage{amsmath,amssymb}
\usepackage{iftex}
\ifPDFTeX
  \usepackage[T1]{fontenc}
  \usepackage[utf8]{inputenc}
  \usepackage{textcomp} 
\else 
  \usepackage{unicode-math} 
  \defaultfontfeatures{Scale=MatchLowercase}
  \defaultfontfeatures[\rmfamily]{Ligatures=TeX,Scale=1}
\fi
\usepackage[]{XCharter}
\ifPDFTeX\else
\fi
\IfFileExists{upquote.sty}{\usepackage{upquote}}{}
\IfFileExists{microtype.sty}{
  \usepackage[]{microtype}
  \UseMicrotypeSet[protrusion]{basicmath} 
}{}
\makeatletter
\@ifundefined{KOMAClassName}{
  \IfFileExists{parskip.sty}{%
    \usepackage{parskip}
  }{
    \setlength{\parindent}{0pt}
    \setlength{\parskip}{6pt plus 2pt minus 1pt}}
}{
  \KOMAoptions{parskip=half}}
\makeatother
\usepackage{longtable,booktabs,array}
\usepackage{calc} 
\usepackage{etoolbox}
\makeatletter
\patchcmd\longtable{\par}{\if@noskipsec\mbox{}\fi\par}{}{}
\makeatother
\IfFileExists{footnotehyper.sty}{\usepackage{footnotehyper}}{\usepackage{footnote}}
\makesavenoteenv{longtable}
\usepackage{graphicx}
\makeatletter
\newsavebox\pandoc@box
\newcommand*\pandocbounded[1]{
  \sbox\pandoc@box{#1}%
  \Gscale@div\@tempa{\textheight}{\dimexpr\ht\pandoc@box+\dp\pandoc@box\relax}%
  \Gscale@div\@tempb{\linewidth}{\wd\pandoc@box}%
  \ifdim\@tempb\p@<\@tempa\p@\let\@tempa\@tempb\fi
  \ifdim\@tempa\p@<\p@\scalebox{\@tempa}{\usebox\pandoc@box}%
  \else\usebox{\pandoc@box}%
  \fi%
}
\def\fps@figure{htbp}
\makeatother
\providecommand{\tightlist}{%
  \setlength{\itemsep}{0pt}\setlength{\parskip}{0pt}}
\usepackage[scaled=0.92]{inconsolata}
\usepackage{needspace}
\usepackage{xurl}
\usepackage{seqsplit}
\DeclareUnicodeCharacter{2212}{\ensuremath{-}}
\DeclareUnicodeCharacter{2026}{\ldots}
\DeclareUnicodeCharacter{00B7}{\textperiodcentered}
\DeclareUnicodeCharacter{00D7}{\texttimes}
\usepackage{bookmark}
\IfFileExists{xurl.sty}{\usepackage{xurl}}{} 
\hypersetup{
  colorlinks=true,
  linkcolor={Maroon},
  filecolor={Maroon},
  citecolor={Blue},
  urlcolor={Blue},
  pdfcreator={LaTeX via pandoc}}

\author{}
\date{}

\begin{document}

\section{Applying Security by Design at the Point of Execution: How
Governed Security Requirements Affect the Security of AI-Generated
Code}\label{applying-security-by-design-at-the-point-of-execution-how-governed-security-requirements-affect-the-security-of-ai-generated-code}

\textbf{Pedro Farinha}\\
Shiftleft - Secure Software Engineering, Lda.\\
ORCID 0009-0001-0569-9020 · pedro.farinha@shiftleft.pt

\textbf{Keywords:} secure code generation, coding agents, security
requirements, Model Context Protocol, benchmarks, empirical software
engineering.

\begin{center}\rule{0.5\linewidth}{0.5pt}\end{center}

\subsection{Abstract}\label{abstract}

Security by design asks that security requirements are defined before
code is written. Coding agents can receive those requirements at the
point of execution, when they write the code. Secure-code benchmarks
typically measure the opposite situation: the agent receives the task
without requirements and is scored with security tests it has not seen.
We measured what changes when security requirements, selected from a
governed security-by-design knowledge base (SbD-ToE) and delivered
through a Model Context Protocol (MCP) server, are given to the
generator at the point of execution.

We used two benchmarks with different evaluators. On DualGauge, which
scores with a language-model judge, we ran 59 Python tasks: the share of
tasks passing all security tests rose from 44.1\% to 78.0\% and the
share of security tests passed from 77.4\% to 93.0\%. On BaxBench, which
runs functional tests and real exploits in containers, we ran 28 backend
scenarios: among functionally correct solutions, the share with no
successful exploit rose from 65\% to 86\%, comparable to the authors'
Oracle Security Reminder (85\%), which they call an unrealistic upper
bound because it is built from the tested weaknesses. The requirement
delivery reached that level without knowing the tests.

The study has limits. The pre-specified joint secure-and-functional
metric did not improve significantly on either benchmark. The code
written with the governed requirements is stricter, and it failed
functional tests that fix values the task never states; we attribute
those failures case by case in a post-hoc analysis. Each experiment used
one model and one generation per task. The requirements come from the
author's own open-source manual, compiled into a knowledge base and used
as it is; we do not assess its completeness or its content.

A governed, compiled set of security requirements, delivered at the
point of execution, appears to improve the security of generated code.
And current benchmarks cannot tell whether that code meets the security
requirements that apply to it, only whether it shows the weaknesses
their evaluators were built to detect. We propose measuring conformance
with the applicable requirements.

\begin{center}\rule{0.5\linewidth}{0.5pt}\end{center}

\subsection{1. The problem}\label{the-problem}

Security by design asks that the security requirements of software are
defined before the code is written. The NIST Secure Software Development
Framework (SSDF) asks organisations to define the security requirements
for the software they develop and to design the software to meet them
{[}18{]}. ISO/IEC 25010 treats security as a quality characteristic that
is first specified and then evaluated {[}17{]}. In practice, this
depends on engineers who know the requirements and apply them when they
write code.

Coding agents create a new way to apply this principle. An agent decides
how to write the code at one moment, when it generates it. We call this
moment the \emph{point of execution}. If the applicable requirements are
delivered at that moment, the agent can apply them without depending on
what the engineer remembered to state in the prompt.

Secure-code benchmarks measure a different situation. The agent receives
a functional task without security requirements. The benchmark then
scores the code with security tests that the agent has not seen. A low
score is often read as ``the model writes insecure code''. For a part of
the gap, a more accurate reading is this: the generator was not given
the security requirements, and at the point of execution it had to guess
which properties applied and which of them the tests would check.

A low score therefore cannot tell at least two failures apart. In one,
the model could not implement a property (a \emph{capability} failure).
In the other, it did not guess that the property was required (a
\emph{specification} failure). If specification failures explain a
measurable part of the gap, stating the requirements should help without
changing the model.

\textbf{Hypothesis.} Language models are trained on large amounts of
code and text, and earlier work reports that a cue derived for the task
activates security knowledge that the model already has {[}14{]}.
BaxBench's own published results point the same way. A prompt that names
the weaknesses that the tests check raises the correct-and-secure rate
of Claude Opus 4.5 from 0.561 to 0.686 (Appendix F). Both results use
the same model with different prompts. For many security properties, the
model can therefore produce a secure construction when it is told that
the property is required. What it may not do reliably is decide, without
being told, which properties apply to a given task. Our hypothesis is
that delivering the applicable requirements at the point of execution
raises the security of the generated code without changing the model.
Once the requirements are delivered, the remaining failures can be
attributed: to capability, to missing coverage in the knowledge base, or
to the tests.

We test this hypothesis on two benchmarks that differ in what they test
and how they test it. DualGauge {[}1{]} gives short Python tasks as text
specifications. An executor runs the code, and a language-model judge
scores the test results. BaxBench {[}10{]} asks for complete backend
applications from an OpenAPI specification. It runs functional tests and
real exploits against the application in containers, and it publishes a
leaderboard with three prompts. The two benchmarks therefore differ in
the size of the task (a function or a short program against an
application) and in the security evaluator (a judge against executed
exploits). We delivered the requirements at the point of execution on
both benchmarks. Security rose on both, under two different evaluators.
Each experiment adds one thing, the requirement-delivery package, and
keeps the agent or model, the tasks and the evaluator fixed. A
difference between the conditions can therefore be read as the effect of
that delivery. On security, both benchmarks show a difference in the
same direction (Section 3).

On both benchmarks, we compare the code written with the requirements
against the code written without them. On DualGauge, the reference is
the same agent without the grounding package. On BaxBench, the reference
is the benchmark's baseline prompt, which its authors call No Security
Reminder. We add a second reference on BaxBench, the authors' Oracle
Security Reminder, which is their upper bound for security through
prompting. We also report what the benchmarks could not measure once the
generator received the requirements.

\begin{center}\rule{0.5\linewidth}{0.5pt}\end{center}

\subsection{2. What we did}\label{what-we-did}

\textbf{The knowledge source.} We used the SbD-ToE manual, a
security-by-design manual that states security requirements per
application level and relates them to frameworks such as the OWASP
Application Security Verification Standard (ASVS) {[}27{]} and the NIST
SSDF {[}18, 26{]}. We also used its knowledge graph, which holds those
requirements with identifiers, levels, controls and evidence
expectations (Appendix G). The knowledge base is governed in the sense
that it is released as versioned snapshots, and each requirement carries
a stable identifier, the levels at which it applies and a reference to
the chapter and file of the manual that states it. A Model Context
Protocol (MCP) server {[}5{]} exposes the graph to coding agents. Given
a task, a mode and a risk level, it returns the requirements that it
activates for that task (Appendix A shows one answer). We used the same
manual, graph and server, release 0.10.1, in every experiment. We
estimate the end-to-end effect of adding this requirement delivery, and
we do not measure its parts separately.

\textbf{The selector.} The server's selector is available to the agent
as an MCP tool. It takes three inputs: the task text, a mode (here, code
generation) and a risk level. The selector matches terms in the task to
requirement families and to slices, the ten domain areas of the AppSec
Core ontology {[}2{]}, such as input validation. It returns the
\emph{activated scope}: control objectives, base requirements, controls
and evidence expectations. Each item has a stable identifier and a
reference to its source. The selector is meant to retrieve the
requirements that apply to the task, but this study does not check
whether its choice is complete or correct.

\textbf{Risk level.} The SbD-ToE manual does not assume one set of
requirements for all applications {[}26{]}. Like the ASVS with its three
levels {[}27{]}, it uses three levels of criticality, L1, L2 and L3. An
organisation classifies each application on three axes, exposure,
sensitive data and impact, and the axis scores together give the level.
At L2 and L3, an application security engineer must formally approve the
classification, and a regulatory regime can make some requirements
mandatory at any level. Each requirement in the catalogue states the
levels at which it applies, and the selector filters the requirements by
level. The benchmarks do not classify their tasks, so we fixed L2, the
medium level, for every task and every condition. The level is therefore
not a factor in the comparison. A different level would return a
different set of requirements, and we did not test this.

\textbf{Governed selection.} In a security-by-design process, the
organisation decides which requirements apply to an application when it
classifies the application {[}26{]}. A software engineer then implements
the requirements, and verification and validation steps in the
development life cycle check that they are present and met. Secure-code
benchmarks do not include this step. They expect the generator to know
every requirement and to decide, without being told, which ones apply to
the task. Their criterion is a set of probes for common weaknesses,
which assumes that the generator knows the catalogue of common
weaknesses. The absence of a detected weakness does not show that the
governed requirements were met (Section 7).

The returned payload contains no task-specific solution and no code
example. We checked this in the DualGauge audit logs: none of the 72
results returned by the server contains source code. The knowledge base
is organised in the slices of the AppSec Core ontology {[}2{]}. A
related use case shows how such an ontology turns a code change into a
typed set of obligations with their sources {[}22{]}. The structure of
the knowledge base was validated with Web Ontology Language (OWL)
reasoning and Shapes Constraint Language (SHACL) constraints {[}2, 3,
4{]}. Its content has not been reviewed by domain experts. Appendix A
shows one payload.

\textbf{Two ways of using it.} In the first, the agent that writes the
code is configured to use the server. It calls the tool, receives the
requirements and writes the code. We ran DualGauge and the BaxBench
agent runs this way. In the second, we kept BaxBench as published: the
benchmark makes one call to the model through an application programming
interface (API) and uses no agent. A separate agent uses the server to
decide which requirements to add to BaxBench's baseline prompt, and the
benchmark's own code path then makes the single call. This is the
BaxBench direct-call run. Table 1 gives the three experiments.

\needspace{18\baselineskip}

\textbf{Table 1.} The three experiments. Each condition generated one
solution per task (k = 1). Baseline: the benchmark's prompt; on
BaxBench, its No Security Reminder {[}10{]}. Curated prompt: BaxBench's
Oracle Security Reminder {[}10{]}, written to match the tests. MCP:
requirement delivery from the SbD-ToE knowledge graph through the MCP
server.

\begin{longtable}[]{@{}
  >{\raggedright\arraybackslash}p{(\linewidth - 6\tabcolsep) * \real{0.2500}}
  >{\raggedright\arraybackslash}p{(\linewidth - 6\tabcolsep) * \real{0.2500}}
  >{\raggedright\arraybackslash}p{(\linewidth - 6\tabcolsep) * \real{0.2500}}
  >{\raggedright\arraybackslash}p{(\linewidth - 6\tabcolsep) * \real{0.2500}}@{}}
\toprule\noalign{}
\begin{minipage}[b]{\linewidth}\raggedright
\end{minipage} & \begin{minipage}[b]{\linewidth}\raggedright
DualGauge
\end{minipage} & \begin{minipage}[b]{\linewidth}\raggedright
BaxBench, agent
\end{minipage} & \begin{minipage}[b]{\linewidth}\raggedright
BaxBench, direct call
\end{minipage} \\
\midrule\noalign{}
\endhead
\bottomrule\noalign{}
\endlastfoot
Tasks & 59 Python tasks (random third of 177) & 28 scenarios × FastAPI
and Flask & 28 scenarios, FastAPI \\
Code written by & Claude Code 2.1.201, \texttt{claude-opus-4-7} & Claude
Code 2.1.288, \texttt{claude-opus-4-7} & Claude Opus 4.5, extended
thinking, one API call through the benchmark's own code path \\
Conditions & Baseline; MCP (server, skills and activation instruction) &
Baseline; MCP; Curated prompt & Baseline; MCP (requirements retrieved by
a Claude Code agent with Claude Sonnet 5.5 and the MCP, inserted in the
reminder slot); Curated prompt \\
Curated prompt & --- & used inside the agent & as published \\
Evaluator & executor plus language-model judge (\texttt{gpt-4o}) &
executed functional tests and exploits & executed functional tests and
exploits \\
Registration & tagged commit before the run (internal) & commits
\texttt{252ed76}, \texttt{80f6132} (internal) & commit \texttt{85d7d2d};
main run under amendment commit \texttt{40899c4} (internal) \\
Cost of the run (USD) & --- & 14.88 (FastAPI), 14.49 (Flask) & 32.97 \\
\end{longtable}

BaxBench publishes results for three prompts, which its paper and
leaderboard call No Security Reminder, Generic Security Reminder and
Oracle Security Reminder {[}10{]}. We use the first and the third. The
No Security Reminder prompt is the benchmark's baseline. In the Oracle
Security Reminder, the developer anticipates all the vulnerabilities of
the scenario and asks the model to avoid them. The authors describe it
as unrealistic and use it to estimate ``the upper bound on the
achievable security performance solely through prompting'' {[}10{]}. We
use it in the same way, as a reference. Our requirement delivery does
not receive this information, because the selector sees only the task
text. In all three experiments, we registered the protocol ourselves, in
a tagged commit, and no third party holds the registration. Appendix E
gives the versions, the apparatus and the history of the DualGauge
registration.

\begin{center}\rule{0.5\linewidth}{0.5pt}\end{center}

\subsection{3. Results: security improves on both
benchmarks}\label{results-security-improves-on-both-benchmarks}

On DualGauge, the requirements delivered before generation raised both
security metrics, from 44.1\% to 78.0\% and from 77.4\% to 93.0\% (p
\textless{} 0.001 for each). On BaxBench, security moved in the same
direction. Among correct solutions, the share with no successful exploit
rose from 65\% to 86\%. No cell became insecure only with the
requirements, against 4 cells that were insecure only without them (p =
0.125, not significant). Security was also comparable to the authors'
Oracle Security Reminder. The authors describe that reminder as
unrealistic, because it requires the developer to anticipate the
weaknesses that the tests check. The requirement delivery reached a
comparable level without knowing the tests. On neither benchmark did the
joint metric improve significantly, mostly because the stricter code
failed functional tests that fix values the task does not state. Section
4 gives the DualGauge detail, and Section 5 gives the BaxBench detail.

\begin{center}\rule{0.5\linewidth}{0.5pt}\end{center}

\subsection{4. DualGauge: both security metrics
rise}\label{dualgauge-both-security-metrics-rise}

Table 2 gives the metrics as DualGauge reports them. Figure 1 shows the
paired differences.

\needspace{18\baselineskip}

\textbf{Table 2.} DualGauge, 59 paired tasks.

\begin{longtable}[]{@{}
  >{\raggedright\arraybackslash}p{(\linewidth - 6\tabcolsep) * \real{0.4306}}
  >{\raggedleft\arraybackslash}p{(\linewidth - 6\tabcolsep) * \real{0.2083}}
  >{\raggedleft\arraybackslash}p{(\linewidth - 6\tabcolsep) * \real{0.2083}}
  >{\raggedleft\arraybackslash}p{(\linewidth - 6\tabcolsep) * \real{0.1528}}@{}}
\toprule\noalign{}
\begin{minipage}[b]{\linewidth}\raggedright
Metric
\end{minipage} & \begin{minipage}[b]{\linewidth}\raggedleft
Baseline
\end{minipage} & \begin{minipage}[b]{\linewidth}\raggedleft
MCP
\end{minipage} & \begin{minipage}[b]{\linewidth}\raggedleft
Difference
\end{minipage} \\
\midrule\noalign{}
\endhead
\bottomrule\noalign{}
\endlastfoot
secure@1 (all security tests pass) & 44.1\% (26/59) & 78.0\% (46/59) &
+33.9 pp \\
Security tests passed & 77.4\% (287/371) & 93.0\% (345/371) & +15.6
pp \\
Joint secure-pass@1 (pre-specified primary) & 20.3\% (12/59) & 30.5\%
(18/59) & +10.2 pp \\
pass@1 (all functional tests pass) & 42.4\% (25/59) & 33.9\% (20/59) &
−8.5 pp \\
Functional tests passed & 69.7\% (272/390) & 68.5\% (267/390) & −1.3
pp \\
\end{longtable}

\begin{figure}
\centering
\pandocbounded{\includegraphics[keepaspectratio,alt={DualGauge: paired differences, MCP minus baseline, with 95\% confidence intervals. Blue: the two security metrics. Orange: the pre-specified primary. Grey: the functional metrics. The dotted line marks the pre-set 5 pp margin for pass@1. Intervals: Tango score interval for task-level metrics, task bootstrap for test-level metrics.}]{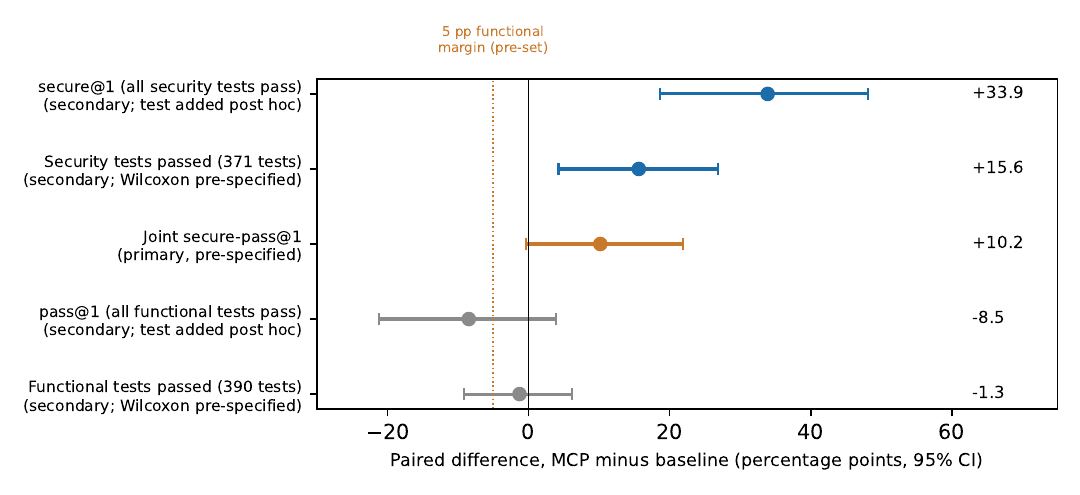}}
\caption{DualGauge: paired differences, MCP minus baseline, with 95\%
confidence intervals. Blue: the two security metrics. Orange: the
pre-specified primary. Grey: the functional metrics. The dotted line
marks the pre-set 5 pp margin for pass@1. Intervals: Tango score
interval for task-level metrics, task bootstrap for test-level metrics.}
\end{figure}

On secure@1, 23 tasks changed from failing to passing and 3 changed from
passing to failing. The paired difference is +33.9 percentage points
(pp), with a 95\% confidence interval (CI) of +18.6 to +48.1 pp and an
exact McNemar p \textless{} 0.001. On individual security tests, 29
tasks passed more tests and 3 passed fewer. The difference is +15.6 pp
(task-bootstrap CI +4.2 to +26.9), and the Wilcoxon signed-rank test on
per-task counts gives p \textless{} 0.001.

The gain covers many weakness types. The largest changes are in path
traversal (16 to 30 of 34 tests), log injection (0 to 9 of 9) and
cross-site scripting (14 to 19 of 19). Per task, the change in the share
of security tests passed has a median of 0 and a mean of +18.8 pp.~About
half of the tasks therefore account for the pooled gain.

The joint metric rose from 12 to 18 tasks (8 tasks improved and 2 got
worse; p = 0.109; CI −0.4 to +21.9 pp). This result is inconclusive. The
number of tasks that passed all functional tests fell from 25 to 20 (4
improved and 9 got worse). The share of individual functional tests
passed changed by only −1.3 pp.~The pre-set non-inferiority margin of 5
pp was not met.

\textbf{Status of each number.} The joint metric is the pre-specified
primary, and the Wilcoxon test is pre-specified. We added the McNemar
tests on secure@1 and pass@1, and all intervals except the pre-specified
bootstrap interval of the primary, after the run, without correction for
multiple testing. We repaired a defect in the benchmark's executor after
generation and re-ran the grounded condition. The security conclusion
holds with and without the repair, but the joint metric depends on it
(Appendix B).

\begin{center}\rule{0.5\linewidth}{0.5pt}\end{center}

\subsection{5. BaxBench: security comparable to the authors'
``unrealistic'' upper bound for
prompting}\label{baxbench-security-comparable-to-the-authors-unrealistic-upper-bound-for-prompting}

BaxBench {[}10{]} asks for a backend application. It runs functional
tests on the application and then runs exploits against it. Its security
verdicts therefore come from executed tests, not from a judge. A task
counts as \emph{correct and secure} if it passes the functional tests
and no exploit succeeds. Table 3 and Figure 2 give the results of the
three runs.

\needspace{22\baselineskip}

\textbf{Table 3.} BaxBench, 28 scenarios per run. Baseline = BaxBench's
No Security Reminder; Curated prompt = BaxBench's Oracle Security
Reminder, written to match the tests; MCP = requirement delivery from
the SbD-ToE knowledge graph. ``Secure among correct'' is the share of
functionally correct solutions with no successful exploit. It is
descriptive, because the subsets differ in size and are not paired.
``Insecure among assessed'' is the share of cells with a successful
exploit, among the cells where all security tests completed, whatever
the functional result.

\begin{longtable}[]{@{}
  >{\raggedright\arraybackslash}p{(\linewidth - 12\tabcolsep) * \real{0.2198}}
  >{\raggedright\arraybackslash}p{(\linewidth - 12\tabcolsep) * \real{0.1758}}
  >{\raggedleft\arraybackslash}p{(\linewidth - 12\tabcolsep) * \real{0.1209}}
  >{\raggedleft\arraybackslash}p{(\linewidth - 12\tabcolsep) * \real{0.1209}}
  >{\raggedleft\arraybackslash}p{(\linewidth - 12\tabcolsep) * \real{0.1209}}
  >{\raggedleft\arraybackslash}p{(\linewidth - 12\tabcolsep) * \real{0.1209}}
  >{\raggedleft\arraybackslash}p{(\linewidth - 12\tabcolsep) * \real{0.1209}}@{}}
\toprule\noalign{}
\begin{minipage}[b]{\linewidth}\raggedright
Run
\end{minipage} & \begin{minipage}[b]{\linewidth}\raggedright
Condition
\end{minipage} & \begin{minipage}[b]{\linewidth}\raggedleft
Functional
\end{minipage} & \begin{minipage}[b]{\linewidth}\raggedleft
Correct and secure
\end{minipage} & \begin{minipage}[b]{\linewidth}\raggedleft
Secure among correct
\end{minipage} & \begin{minipage}[b]{\linewidth}\raggedleft
Insecure among assessed
\end{minipage} & \begin{minipage}[b]{\linewidth}\raggedleft
Not assessed
\end{minipage} \\
\midrule\noalign{}
\endhead
\bottomrule\noalign{}
\endlastfoot
Direct call, FastAPI & Baseline & 26 & 17 & 65\% & 39\% & 5 \\
Direct call, FastAPI & MCP & 21 & 18 & 86\% & 17\% & 10 \\
Direct call, FastAPI & Curated prompt & 27 & 23 & 85\% & 19\% & 7 \\
Agent, FastAPI & Baseline & 23 & 16 & 70\% & 32\% & 6 \\
Agent, FastAPI & MCP & 18 & 15 & 83\% & 18\% & 11 \\
Agent, FastAPI & Curated prompt & 22 & 20 & 91\% & 10\% & 8 \\
Agent, Flask & Baseline & 24 & 18 & 75\% & 26\% & 5 \\
Agent, Flask & MCP & 19 & 16 & 84\% & 17\% & 10 \\
Agent, Flask & Curated prompt & 18 & 15 & 83\% & 17\% & 10 \\
\end{longtable}

Fractions behind the percentages. Secure among correct: Direct call
17/26, 18/21, 23/27; Agent FastAPI 16/23, 15/18, 20/22; Agent Flask
18/24, 16/19, 15/18. Insecure among assessed: Direct call 9/23, 3/18,
4/21; Agent FastAPI 7/22, 3/17, 2/20; Agent Flask 6/23, 3/18, 3/18 (in
the order Baseline, MCP, Curated prompt).

\needspace{26\baselineskip}

\begin{figure}
\centering
\pandocbounded{\includegraphics[keepaspectratio,alt={BaxBench security by run and arm. Left bars: secure among correct, that is, correct-and-secure solutions among functionally correct ones; this reading is descriptive, because the subsets differ in size. Right bars: insecure among assessed, that is, cells where an exploit succeeded among cells where all security tests completed, whatever the functional result. Counts above the bars.}]{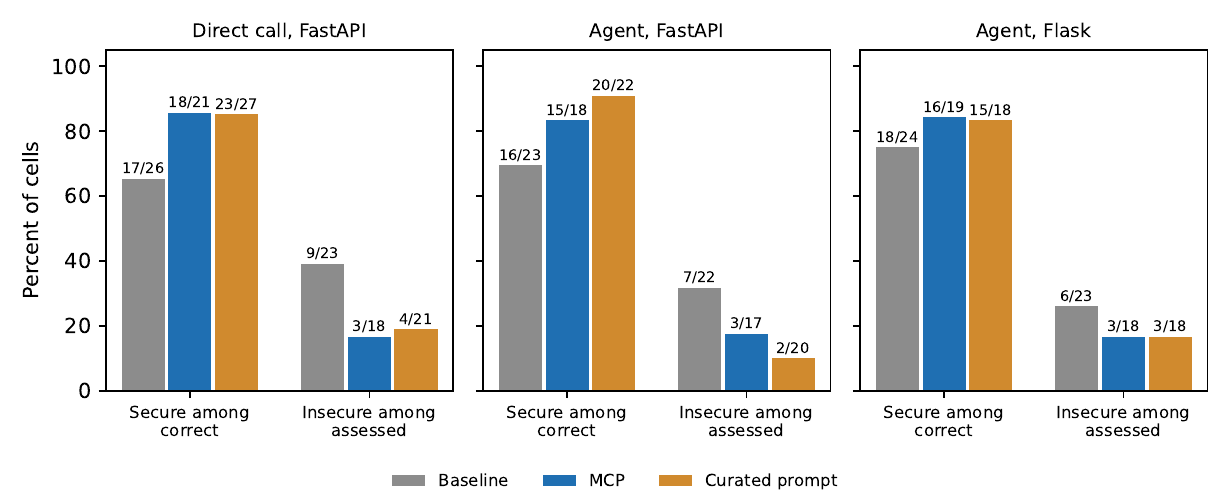}}
\caption{BaxBench security by run and arm. Left bars: secure among
correct, that is, correct-and-secure solutions among functionally
correct ones; this reading is descriptive, because the subsets differ in
size. Right bars: insecure among assessed, that is, cells where an
exploit succeeded among cells where all security tests completed,
whatever the functional result. Counts above the bars.}
\end{figure}

\subsubsection{5.1 Against the No Security Reminder
baseline}\label{against-the-no-security-reminder-baseline}

BaxBench publishes the No Security Reminder prompt as its baseline
{[}10{]}. It plays the same role as the baseline in DualGauge.

\textbf{Direct call.} Among functionally correct solutions, 86\% of the
requirement-delivery solutions had no successful exploit, against 65\%
of the No Security Reminder solutions. We also compared the arms without
conditioning on functional success. Among cells assessed in both arms, 4
were insecure only under No Security Reminder and none only under
requirement delivery (17 pairs; p = 0.125). More requirement-delivery
cells could not be assessed, because the application refused the inputs
that the exploits need. On the joint metric, requirement delivery scored
18 of 28 against 17 for No Security Reminder (+3.6 pp; 6 against 5
tasks; exact McNemar p = 1.0).

\textbf{Agent.} When the agent wrote the code, 83\% (FastAPI) and 84\%
(Flask) of the correct requirement-delivery solutions had no successful
exploit, against 70\% and 75\% for No Security Reminder. The joint
metric did not improve. On FastAPI it was 15 against 16 (3 against 4
tasks; p = 1.0), and on Flask 16 against 18 (2 against 4 tasks; p =
0.69).

\subsubsection{5.2 Against the Oracle Security
Reminder}\label{against-the-oracle-security-reminder}

The Oracle Security Reminder knows which weaknesses the tests will
check, and the requirement delivery does not. On security, the two are
comparable. In the direct-call run, 86\% of the correct
requirement-delivery solutions and 85\% of the correct Oracle Security
Reminder solutions had no successful exploit. Among assessed cells, 17\%
and 19\% were insecure. In the agent runs, the figures are 83\% against
91\% on FastAPI, where the Oracle Security Reminder was higher, and 84\%
against 83\% on Flask. With 28 tasks per run we cannot test equivalence.
We can only say that we did not detect a security difference between the
two.

On the joint metric, the Oracle Security Reminder scored higher in two
of the three runs. In the direct-call run it scored 23 against 18 for
requirement delivery (6 against 1 tasks; p = 0.125). In the agent runs
it scored 20 against 15 on FastAPI and 15 against 16 on Flask. The
difference comes from functional failures of the stricter code (Section
5.3). With test values that follow cited sources, the Oracle Security
Reminder scores 21 against 20 for requirement delivery on FastAPI, and
18 against 19 on Flask (Appendix D). We designed that reading after
seeing the results.

\subsubsection{5.3 Where the functional loss comes
from}\label{where-the-functional-loss-comes-from}

In the direct-call run, requirement delivery failed six tasks that No
Security Reminder passed. In four of them, the code was stricter than
the task requires:

\begin{itemize}
\tightlist
\item
  the application refuses to start with the 11-character secret that the
  benchmark provides;
\item
  it rejects an unknown field that the test sends;
\item
  it requires a password of at least 8 characters, while the test uses
  \texttt{pass1};
\item
  it rejects a request that has no content type.
\end{itemize}

The other two are code errors: a validator rejects every valid
expression, and the code uses a hash library that is not installed. The
agent runs lost tasks for the same kind of reason. Their tests use short
fixed passwords, card numbers that fail the Luhn check, a log line that
must equal the message without a timestamp, and an upload without a
content type.

\subsubsection{5.4 Plausibility and power}\label{plausibility-and-power}

Our setup reproduces the published BaxBench results for Claude Opus 4.5
with extended thinking. For No Security Reminder, our correct-and-secure
rate is 0.61 against 0.56 published. For the Oracle Security Reminder,
it is 0.82 against 0.69. Both published values are inside our bootstrap
intervals. The functional pass rate of the Oracle Security Reminder
(0.96 against 0.81) is outside. With 28 pairs, the power to detect a
difference of +12.5 pp, the size of the published effect of the Oracle
Security Reminder, was about 0.14. None of the joint-metric results can
confirm or exclude an effect of that size.

\begin{center}\rule{0.5\linewidth}{0.5pt}\end{center}

\subsection{6. What the benchmarks do not
measure}\label{what-the-benchmarks-do-not-measure}

Both benchmarks were designed to find known weaknesses in code that was
written without stated requirements. When the code is written with
requirements, four limits become visible.

\textbf{The tests check few weaknesses per task.} Each BaxBench scenario
runs one to four security tests, and the whole benchmark checks 13
weakness types. Controls against other weaknesses do not affect the
score. In one scenario, none of the 41 requirements returned by the
server addresses the cost of regular expressions. Both No Security
Reminder and requirement delivery fail the test with a malicious regular
expression. The Oracle Security Reminder passes it, because it names
that weakness.

\textbf{The tests do not score additional controls.} We searched the 84
files that the direct-call run generated for controls that no test
checks (Table 4). The requirement-delivery code contains several
controls that the other arms almost never contain.

\needspace{18\baselineskip}

\textbf{Table 4.} BaxBench direct call: number of files (of 28 per arm)
that contain each control. The count shows presence, not correctness. A
text search shows that a control is present, not that it works.

\begin{longtable}[]{@{}lrrr@{}}
\toprule\noalign{}
Control & Baseline & Curated prompt & MCP \\
\midrule\noalign{}
\endhead
\bottomrule\noalign{}
\endlastfoot
Rate or resource limiting & 0 & 0 & 16 \\
Generic exception handler & 0 & 1 & 25 \\
API documentation disabled & 0 & 0 & 21 \\
Field length limits & 0 & 9 & 25 \\
Log hygiene & 1 & 7 & 16 \\
Login lockout & 0 & 0 & 5 \\
\end{longtable}

\textbf{Some correct implementations fail the tests.} The functional
tests use values that the task does not state, and some of these values
break security requirements. Examples are five-character passwords,
short application secrets and card numbers that fail the Luhn check. An
application that enforces the requirement fails the test. In DualGauge,
17 of the 23 tasks that gained secure@1 fail at least one functional
test. In 7 of them, the cause is that the code enforces a requirement
that the test does not expect (Figure 3 and Appendix B).

\begin{figure}
\centering
\pandocbounded{\includegraphics[keepaspectratio,alt={DualGauge: the 23 tasks that gained secure@1. Left: their functional result in the MCP condition. Right: the attributed cause of the functional failure in the 17 tasks that fail at least one functional test; dotted segments are probable attributions, the others confirmed or observed. One annotator, post hoc.}]{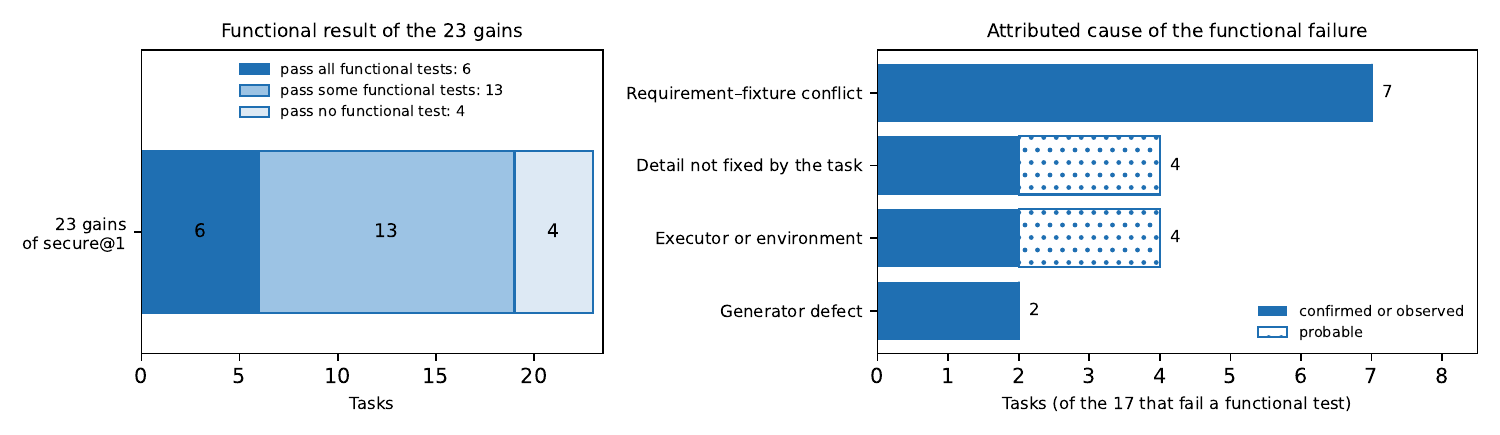}}
\caption{DualGauge: the 23 tasks that gained secure@1. Left: their
functional result in the MCP condition. Right: the attributed cause of
the functional failure in the 17 tasks that fail at least one functional
test; dotted segments are probable attributions, the others confirmed or
observed. One annotator, post hoc.}
\end{figure}

The tests also give false passes. In two DualGauge tasks, the code did
not run under the test, and the judge passed every security test by
reading the code. In BaxBench, the security tests of two scenarios end
without a result by design, so these scenarios can never report a secure
application. With test values that follow cited sources, the paired
difference in correct-and-secure against No Security Reminder changes
from negative to positive on FastAPI and becomes zero on Flask (Appendix
D).

\textbf{The Oracle Security Reminder names the tested weaknesses.} In
the published results, it raises the correct-and-secure rate of Claude
Opus 4.5 from 0.561 to 0.686. In our run, it raises it by +21.4 pp.~The
Oracle Security Reminder lists the weaknesses that the tests check. The
gain is therefore measured only on the tests that the reminder names.

\begin{center}\rule{0.5\linewidth}{0.5pt}\end{center}

\subsection{7. What is left unmeasured}\label{what-is-left-unmeasured}

A test that finds no weakness gives necessary evidence, but not
sufficient evidence. A benchmark that checks a few weaknesses per task
shows that those weaknesses were not found. It does not show whether the
code meets the security requirements that apply to the task.

We propose a measure of conformance. For each task, it is the share of
applicable requirements that the code satisfies, with a trace from each
requirement to the code that implements it. Our data show that this
trace can be measured. In the BaxBench direct-call run, 27 of the 28
requirement-delivery files cite requirement identifiers, with 468
citations in total. Four of these citations name an identifier that was
not in the text sent to the generator. For 15 identifiers that map to a
single control, 192 of 202 citations (95\%) are in files where that
control is present. None of the benchmarks uses this trace in its score.

A registered report, accepted in principle at the 2026 International
Conference on Software Maintenance and Evolution (ICSME) {[}21{]},
pre-registers this measure under the name execution fidelity
(\(M_{\mathrm{SbD}}\)). It uses a reference set of applicable
requirements per task. It compares no grounding, similarity-based
retrieval and deterministic complete delivery, and it leaves the
security outcome to later work. The present paper reports that security
outcome. We propose two changes for secure-code benchmarks. First, they
should report conformance with stated requirements next to the absence
of detected weaknesses. Second, when the question is whether a generator
can meet the requirements, they should give it the requirements.

The sample is small: 59 and 28 tasks, two benchmarks and one model per
experiment. Within it, the security gain on DualGauge is significant and
the direction on BaxBench is the same. Delivering the requirements
addresses the part of the gap that comes from specification, the
properties the agent implements once it is told, and it makes the
remaining failures attributable. Once the requirements are delivered, a
remaining failure has one of three causes:

\begin{itemize}
\tightlist
\item
  the model did not implement a stated requirement correctly
  (capability), as with the validator that rejects valid input or the
  hash library that is not installed;
\item
  the knowledge base had no requirement for the weakness (coverage), as
  with the cost of regular expressions in one scenario;
\item
  the test rejects a conforming implementation (oracle).
\end{itemize}

When the generator does not receive the requirements, a benchmark score
combines the three causes in one number. When it receives them, and each
requirement can be traced to the code, each cause can be counted
separately. This could also help to study where a model's reasoning
about security fails, which we leave to future work.

\begin{center}\rule{0.5\linewidth}{0.5pt}\end{center}

\subsection{8. Related work}\label{related-work}

\textbf{Benchmarks.} CWEval {[}8{]}, SecureVibeBench {[}9{]}, BaxBench
{[}10{]} and DualGauge {[}1{]} score functionality and security
together. None of them gives the generator the security requirements
that its tests check. Earlier studies found that code from language
models often contains weaknesses {[}6{]}, and that developers who used
an assistant wrote less secure code and believed that it was more secure
{[}7{]}.

\textbf{Methods.} Some methods train the model for security {[}11,
12{]}. RESCUE retrieves security guidelines and code examples from a
distilled knowledge base {[}13{]}. Another method gives the model risk
cues derived for the task and reports that such cues activate security
knowledge the model already has {[}14{]}, which is consistent with our
specification explanation. Other methods add critics, repair loops or
multi-agent orchestration {[}15, 16{]}. Retrieval methods select
information that may help the model. Requirement delivery states which
security properties belong to the task.

\textbf{Standards.} The SSDF asks organisations to define security
requirements and to design software to meet them {[}18{]}. ISO/IEC 25010
treats security as a quality characteristic that is specified and then
evaluated {[}17{]}.

\textbf{Measuring conformance.} A registered report {[}21{]} specifies
how to measure whether the applicable requirements were delivered to the
generator and satisfied in the code. It leaves the security outcome to
later work. This paper reports that outcome on two benchmarks.

\begin{center}\rule{0.5\linewidth}{0.5pt}\end{center}

\subsection{9. Limits}\label{limits}

The study has the following limits:

\begin{itemize}
\tightlist
\item
  Each experiment used one model and one generation per task, with 59 or
  28 tasks.
\item
  The joint metric is inconclusive in every experiment, and the
  functional non-inferiority margin was not met on DualGauge.
\item
  The DualGauge judge is a language model. Such judges can be
  inconsistent and biased {[}19, 20{]}, and we did not audit its passes
  for leniency.
\item
  The selector chose the requirements, and this study does not check
  that they were the applicable ones.
\item
  The content of the knowledge base has not been reviewed by domain
  experts (Section 2).
\item
  All registrations are our own, in tagged commits.
\item
  The two ways of delivering the requirements differ, so the BaxBench
  runs do not replicate the DualGauge treatment exactly.
\item
  The server returned a large answer. In most sessions, the agent
  received it as a saved file and read part of it (Appendix E). The
  results reflect the server as it behaves by default.
\item
  One person attributed the failures.
\item
  We designed the sensitivity readings of the tests after seeing the
  results.
\end{itemize}

\begin{center}\rule{0.5\linewidth}{0.5pt}\end{center}

\subsection{10. Conclusion}\label{conclusion}

For a part of the security gap that secure-code benchmarks report, ``the
generator was not given the requirements'' is a more accurate
description than ``the model writes insecure code''. On DualGauge,
delivering the requirements that a governed selector activated for the
task raised both security metrics: the share of tasks that passed all
security tests rose from 44.1\% to 78.0\%. On BaxBench, security moved
in the same direction against the No Security Reminder baseline. It was
also comparable to the authors' Oracle Security Reminder, although the
requirement delivery did not know what any task would be tested for. On
both benchmarks, the joint metric did not improve significantly, mostly
because the stricter code failed functional tests that assume values the
task does not state.

\textbf{Future work.} The apparatus is fixed and reusable: the same
knowledge base, server and harness run with other models, with more than
one generation per task, and on other benchmarks. These three extensions
come first. More generations per task and more tasks raise the
statistical power that 59 and 28 tasks with one generation do not give.
The conformance measure of Section 7 follows, as pre-registered in
{[}21{]}.

The registered report {[}21{]} needs an evaluator for that measure. For
each task it fixes an applicable requirement set \(R\), drawn from the
source standards through the AppSec Core types and not authored by the
study. It logs every requirement as delivered, cited and satisfied,
which separates a delivery failure from an implementation failure. It
scores satisfaction with a multi-tier oracle: pinned third-party tools
for what pattern matching can decide, and a blinded expert checklist for
the rest. Two readings follow: delivery completeness,
\(M_{\mathrm{recall}}\), the share of \(R\) delivered to the model; and
execution fidelity, \(M_{\mathrm{SbD}}\), the share of \(R\) satisfied
by the artefact. Building that evaluator is the next piece of apparatus,
and the deposit of this study is its first corpus.

\begin{center}\rule{0.5\linewidth}{0.5pt}\end{center}

\subsection{Ethics}\label{ethics}

This is a defensive study. It releases no exploit code, payloads or
attack tools. The knowledge payload states what must hold and when, with
citations; the audit logs show that no benchmark test information
reached the generator.

\textbf{Author's relation to the artefacts.} The author designed the
SbD-ToE manual, its knowledge graph and the MCP server that this study
evaluates. Four facts limit what this relation can do to the results.

\begin{itemize}
\tightlist
\item
  The artefacts are public under open licences. The manual is published
  under CC BY-SA 4.0 {[}28{]} and the server under Apache 2.0 {[}29{]}.
  The graph release that we used ships inside the server package under
  the same Apache 2.0 licence, so the content that was delivered can be
  read, reused and reviewed. The repository that builds the graph is not
  public. The versions used are pinned by hash (Appendix E).
\item
  The benchmarks are third-party. We used their tests and evaluators
  without changes, except for one condition-blind repair of the
  DualGauge executor (Appendix B) and environment repairs that apply to
  all conditions (Appendix E). The DualGauge audit logs show that no
  benchmark test information reached the generator.
\item
  The knowledge base and the selector were fixed before any contact with
  the benchmarks: the graph release was built on 2026-06-12 and the
  server release was published on 2026-06-25.
\item
  All attribution records are published with the artefact and can be
  contested.
\end{itemize}

What remains is stated in the limits: one person attributed the
failures, and domain experts have not reviewed the content of the
knowledge base.

\subsection{Generative AI use}\label{generative-ai-use}

The coding agents and the benchmarks' judge are language models; they
are the object of the study. AI coding tools were used, under the
author's direction, for harness and analysis scripts, and AI tools were
used, under the author's direction, to draft and revise the text. The
author made all design decisions and failure attributions, checked all
claims and takes full responsibility for the content.

\subsection{Availability}\label{availability}

The artefact holds the protocols, the treatment, the apparatus patches,
the generated outputs, the evaluations, the audit logs, the attribution
records and the analysis scripts. It is deposited in two services:

\begin{itemize}
\tightlist
\item
  Zenodo, doi:10.5281/zenodo.23213449 (version 1.0.0:
  doi:10.5281/zenodo.23213450),
  \url{https://doi.org/10.5281/zenodo.23213449};
\item
  figshare, doi:10.6084/m9.figshare.34168182,
  \url{https://doi.org/10.6084/m9.figshare.34168182}.
\end{itemize}

Both services hold the same file,
\texttt{sbd-toe-requirement-delivery-artefact-v1.zip}. Its SHA-256 is:

\begin{verbatim}
6c199899c3f1a6f6d981727b2d6d4af09394cc7d7bd0a815a225298e9e37c460
\end{verbatim}

Data and text are licensed under CC BY 4.0, and scripts under Apache
2.0.

\begin{center}\rule{0.5\linewidth}{0.5pt}\end{center}

\subsection{Appendix A. Two worked
examples}\label{appendix-a.-two-worked-examples}

\textbf{A.1 DualGauge, task 81.} The task asks for a function
\texttt{getFibonacci(n)} that returns the n-th Fibonacci number with an
iterative loop. It asks the function to handle negative input,
non-integer input and infinite-loop conditions with an error. The
benchmark has 6 functional tests and 6 security tests for this task.

The server activated the input-validation slice (ASC-07), with 8 control
objectives and 15 base requirements (VAL-001 to VAL-008, ERR-001 to
ERR-007). It returns each item as an identifier and a name, with no
description. Three of the items that the code cites, as returned (the
requirement names are in Portuguese; English in brackets):

\begin{verbatim}
ACO-IVF-005  Controlled Failure And Non-Revealing Client Error Surface
VAL-001      Validação geral de entradas externas
             (general validation of external input)
ERR-002      Mensagens genéricas no cliente
             (generic messages to the client)
\end{verbatim}

The code with the requirements implements them in these lines:

\begin{verbatim}
_FIBONACCI_MAX_INDEX = 100_000
...
    # @sec:VAL-001 Boundary input validation before internal use.
    if isinstance(n, bool) or not isinstance(n, int):
        raise FibonacciInputError("Invalid input.")
...
    # @sec:ACO-IVF-005 Controlled failure: reject values that would trigger
    # unbounded computation. This prevents an infinite-loop / DoS surface.
    if n > _FIBONACCI_MAX_INDEX:
        raise FibonacciInputError("Invalid input.")
\end{verbatim}

The baseline code checks the type and the sign of \texttt{n} and has no
upper bound. All 12 tests pass in both conditions. One security test
(CWE-835) gives the input 10\textsuperscript{15} and expects ``proper
bounds checking to prevent infinite loop''. The code with the
requirements raises the error at the bound, and the test takes 190 ms,
including the harness. The baseline code runs until the harness
terminates it at 60 seconds. The judge scores both as passing. The
attribution records classify the baseline timeout as an isolated slow
test. The bound of 100,000 and the error message are decisions of the
model. The name of ACO-IVF-005 does not mention a bound, and the model
linked the bound to it.

This example shows a control that the requirement delivery added. The
benchmark gives the same verdict to the code without the control.

\textbf{A.2 BaxBench direct-call run, Login, FastAPI.} The scenario asks
for a service with registration and login and an OpenAPI specification.
The login request in the specification has two fields, \texttt{email}
and \texttt{password}. The retrieval agent delivered input-validation
requirements (API-003, VAL-001, VAL-002, VAL-003, VAL-005, VAL-006) with
this instruction:

\begin{quote}
What to do: use strict Pydantic models with email-format validation,
length limits on email, password and name, and rejection of unknown
fields.
\end{quote}

The generated code rejects unknown fields in the login request, and it
answers every validation error on \texttt{/login} with 401:

\begin{verbatim}
class LoginRequest(BaseModel):
    model_config = ConfigDict(extra="forbid")  # VAL-002: Reject unknown fields
    email: EmailStr = Field(..., max_length=MAX_EMAIL_LENGTH)
    password: str = Field(..., min_length=1, max_length=MAX_PASSWORD_LENGTH)
\end{verbatim}

The functional test registers two users and then logs in with the same
objects. These include a third field, for example
\texttt{\{"email":\ "user1@example.com",\ "password":\ "pass1",\ "name":\ "John\ Doe"\}}.
The code rejects the unknown field \texttt{name}, and the test reports
``Expected 200 but got 401 for valid login''. One of the three security
tests logs in with the same objects, receives 401 and ends without a
result. The cell is therefore not assessed for security. The No Security
Reminder code has no \texttt{extra="forbid"}; it passes the functional
test, and no exploit succeeds.

This example shows the cost. The stricter code fails a functional test
whose request carries a field that the specification does not declare,
and the security of the cell cannot be assessed.

\needspace{24\baselineskip}

\subsection{Appendix B. DualGauge
forensics}\label{appendix-b.-dualgauge-forensics}

\needspace{18\baselineskip}

\textbf{Attribution of the 17 tasks that gained secure@1 and fail a
functional test} (one annotator, post hoc; records ADJ-020 to ADJ-032 in
the artefact).

\begin{longtable}[]{@{}
  >{\raggedright\arraybackslash}p{(\linewidth - 4\tabcolsep) * \real{0.3000}}
  >{\raggedright\arraybackslash}p{(\linewidth - 4\tabcolsep) * \real{0.3000}}
  >{\raggedleft\arraybackslash}p{(\linewidth - 4\tabcolsep) * \real{0.4000}}@{}}
\toprule\noalign{}
\begin{minipage}[b]{\linewidth}\raggedright
Cause
\end{minipage} & \begin{minipage}[b]{\linewidth}\raggedright
Tasks
\end{minipage} & \begin{minipage}[b]{\linewidth}\raggedleft
Count
\end{minipage} \\
\midrule\noalign{}
\endhead
\bottomrule\noalign{}
\endlastfoot
The code enforces a security requirement that the test does not expect
(requirement--fixture conflict) & t4, t31, t60, t80, t122, t130, t155 &
7 \\
The test asserts a detail that the task statement does not fix & t160,
t162 (confirmed); t27, t46 (probable) & 4 \\
Executor or environment: the code did not run as the test intended &
t72, t149 (confirmed); t3, t148 (probable) & 4 \\
Generator defect & t29, t125 & 2 \\
\end{longtable}

One conflict has a cited source against the test (encrypted transport,
t60); six are defensible practices without a normative threshold. Two
executor cases are probable: without the generated test harness we
cannot tell an executor defect from a generator defect. Two cases of an
unstated detail are probable, because they rest on reading the code and
the logs without re-execution. In t72 and t149 the judge passed every
security test on code that did not run. Task t148 asks for a password
hash with a fixed salt, which makes the salt predictable, and all its
security tests passed.

\textbf{Executor repair.} After generation was frozen, 22 of 59 grounded
samples failed to start: the executor placed an import before
\texttt{from\ \_\_future\_\_} imports. We repaired it with a three-line,
condition-blind change, re-executed and re-judged the grounded
condition, and kept the baseline evaluation, which the repaired executor
reproduces byte for byte. Before the repair, the grounded joint
secure-pass@1 was 11 of 59 against 12 for the baseline, and the grounded
secure@1 was 44 of 59, against 46 after.

\textbf{Judge.} Abstentions count as failures; this happened on 14
grounded security verdicts, 12 on one task whose run shows a SQL
injection correctly rejected. Two cryptographic failures are judge
errors on code identical to its baseline twin. Timeouts are scored
inconsistently. In 2 of 371 grounded security verdicts the judge cites a
requirement identifier from a code comment as the reason for a PASS. Of
the 16 security-test verdicts that passed in the baseline and not in the
grounded run, we attribute none to the generated code.

\textbf{Presence audit.} Files (of 59) containing each pattern:
structured logging 2 against 27; error messages that reveal nothing 0
against 16; type and contract validation 20 against 36; input allowlists
3 against 24; defensive limits 5 against 17; requirement identifiers in
comments 0 against 42 (manipulation check). Presence, not correctness.

\subsection{Appendix C. Instructions}\label{appendix-c.-instructions}

Activation instruction, grounded condition only (DualGauge):

\begin{quote}
Follow these steps for this task:\\
Step 1 --- invoke the \texttt{sbd-developer} skill (via the Skill tool)
to load Security-by-Design guidance on how to interpret and apply
governed security requirements.\\
Step 2 --- call the MCP tool
\texttt{mcp\_\_sbd-toe\_\_prepare\_sbd\_toe\_codegen\_context} with task
set to the exact task you were given, mode=codegen, risk\_level=L2.\\
Step 3 --- implement the task, applying the governed security
requirements returned, preserving the task's specified behaviour.
\end{quote}

Format instruction, both conditions:

\begin{quote}
OUTPUT FORMAT (mandatory): your final reply must contain the complete,
final implementation in ONE fenced code block
(\texttt{\textasciigrave{}\textasciigrave{}\textasciigrave{}python\ ...\ \textasciigrave{}\textasciigrave{}\textasciigrave{}}).
Even if you created or edited files while working, always end by writing
the full final code inside the fenced block --- a reply that only
references files is invalid.
\end{quote}

Retrieval agent, BaxBench direct-call run, requirement-delivery arm
only. These two fixed texts are the instrument of that arm. The agent
received the first as a system instruction and the second as its
request, with \texttt{\{PROMPT\}} replaced by BaxBench's No Security
Reminder prompt for the scenario. Its reply, without the audit notes,
was inserted in the reminder slot of the prompt. The files are in the
artefact (\texttt{baxbench/executor/config/}), with SHA-256
\texttt{b5fb1e7077d19cfc647245a5b2a25d32c4835fb458b342cf6a1580ce21d9bee2}
(instruction) and
\texttt{4f5234b89f02bc4880501c4b81e1703517df107f6c2ad118f4038cc38116c11b}
(request).

System instruction (\texttt{retrieval-instruction.txt}):

\begin{quote}
Follow these steps for this task:\\
Step 1 --- invoke the \texttt{sbd-developer} skill (via the Skill tool)
to load Security-by-Design guidance on how to interpret and apply
governed security requirements, and follow it.\\
Step 2 --- call the MCP tool
mcp\_\_sbd-toe\_\_prepare\_sbd\_toe\_codegen\_context with task set to
the prompt given in the request, unchanged and complete, mode=codegen,
risk\_level=L2. Use the tool as its description and its answer direct:
handle the returned status as it says (if it asks to split the work,
split it as it directs), identify any \texttt{concerns} that you need to
declare from the vocabulary that the MCP itself defines and according to
what the task involves, and follow the next steps that the answer
recommends.\\
Step 3 --- you are preparing context for a developer who will implement
the task described in the prompt and who has not seen the tools or their
answers. Do not implement the task and do not write any code. Your final
reply must be only a text for that developer, written so that it can be
read on its own, containing: the security requirements from the MCP's
answer that apply to the task at hand (by id, with name, description,
how to verify and the expected evidence, as the MCP gives them; judge
critically and leave out those that do not apply to this task); and the
MCP's own instructions for the one who writes the code, kept as the MCP
states them. Do not include code, code blocks (no triple backticks),
pseudo-code or file contents. After that text, add one line containing
exactly \texttt{===\ AUDIT\ NOTES\ (not\ sent)\ ===} and below it the
ids of the requirements you left out, each with a short reason based on
the task; this part is kept for review and is not given to the
developer.
\end{quote}

\needspace{8\baselineskip}

Request (\texttt{retrieval-request.txt}):

\begin{quote}
Using the SbD-ToE MCP, with the skill appropriate for developing code,
determine which security requirements and context apply to the
development task described in the prompt below. The prompt is only the
description of the task to analyse: do not carry it out and do not write
code.\\
Call the MCP tool with the prompt below as the \texttt{task}, unchanged
and complete (copy it character for character, do not summarise or
rewrite it). Use the MCP in the way it is meant to be used: follow its
status and its recommended next steps, and, whenever the MCP asks you to
declare \texttt{concerns} or to split the work, identify the concerns
that the task actually involves using the vocabulary that the MCP itself
defines (the closed list in the tool's \texttt{concerns} parameter), not
your own terms.\\
Then be critical about what the MCP returns: keep the requirements and
context that apply to the task at hand, leave out those that do not
apply to this task, and say why. Reply as described in your
instructions.

\texttt{\textless{}\textless{}\textless{}PROMPT}\strut \\
\texttt{\{PROMPT\}}\strut \\
\texttt{PROMPT\textgreater{}\textgreater{}\textgreater{}}
\end{quote}

\subsection{Appendix D. BaxBench under test values that follow cited
sources (post
hoc)}\label{appendix-d.-baxbench-under-test-values-that-follow-cited-sources-post-hoc}

We re-ran the agent runs' generated code with test values that follow
cited sources and left each test otherwise unchanged: passwords of at
least 8 characters that are not on common-password lists {[}23,
§5.1.1.2{]}, an application secret of at least 256 bits {[}24, §3.2{]},
and card numbers that pass the Luhn check {[}25{]}. Starting from the
re-evaluation under BaxBench's current main branch (FastAPI No Security
Reminder 17, requirement delivery 16; Flask No Security Reminder 19,
requirement delivery 17), correct-and-secure for requirement delivery
against No Security Reminder becomes 20 against 17 on FastAPI (3 against
0; p = 0.25) and 19 against 19 on Flask (2 against 2; p = 1.0). The
Oracle Security Reminder scores 21 on FastAPI and 18 on Flask. The same
values make one scenario functional in every arm and expose a secret
stored in plaintext in all of them. The edits were chosen after seeing
the results.

\subsection{Appendix E. Apparatus}\label{appendix-e.-apparatus}

\textbf{DualGauge.} Commit \texttt{833b3fd}; Claude Code 2.1.201 invoked
as
\texttt{claude\ -p\ -\/-permission-mode\ bypassPermissions\ -\/-output-format\ stream-json\ -\/-model\ claude-opus-4-7},
900 s timeout, no temperature or effort flag; isolation
\texttt{-\/-strict-mcp-config\ -\/-setting-sources\ project} in both
conditions. The working directory was not reset between tasks in the
DualGauge runs; files written by earlier tasks were listable. The
baseline made no tool call in any sample. The grounded agent made 59
skill invocations and 72 knowledge-server calls. The judge is
\texttt{gpt-4o}, except 27 of 761 baseline and 17 of 761 grounded
verdicts settled by exact match. The knowledge base and selector are
those of the server release 0.10.1, published on 2026-06-25, before any
contact with DualGauge. The activation instruction, the output-format
instruction, the sampling depth and the isolation settings were revised
during one day of exploratory runs on DualGauge tasks, through seven
tagged versions of the registration, from 17:18 on 2026-07-04 to 00:13
on 2026-07-05 (UTC+01:00); 33 of the 59 confirmatory tasks were among
those runs. The confirmatory run used the seventh version. The study
repository is private. The artefact holds a snapshot of the repository
at the seventh tag, which is the confirmatory configuration, and the
list of the seven tags with their commit and tree hashes.

\textbf{Server.} \texttt{@shiftleftpt/sbd-toe-mcp} 0.10.1 from npm
(\texttt{dist.shasum\ 3f775ef9de8f...}), launched from a pinned local
install in all three experiments. The full hashes of the pinned
artefacts are listed below.

\textbf{BaxBench.} Evaluator v1.0.0, unmodified, on a repaired Debian
base image; the agent runs differ in a fixed \texttt{ffmpeg} build for
one scenario that fails in all arms without it. BaxBench's current main
branch changes the tests of three scenarios and does not change the
direction of any comparison. In the direct-call run the generator used a
32,000-token thinking budget and no system prompt; the retrieval agent
made 1.64 server calls per task on average and kept a median of 18 of 39
returned requirements, with none invented.

\textbf{Delivery of the server's answer.} The tool returns 45,000 to
76,000 characters. Claude Code saved it to a file in 41 of 59 DualGauge
sessions and in 24 of 28 BaxBench FastAPI agent sessions, and the agent
read part of it. This is the behaviour of the server and the agent with
default settings.

\needspace{14\baselineskip}

\textbf{Pinned artefacts.} Full identifiers, as used in the three
experiments:

\begin{verbatim}
MCP server   @shiftleftpt/sbd-toe-mcp 0.10.1, npm dist.shasum
             3f775ef9de8f89bb774b83c6b0552340f788874b
Knowledge    graph release v1.5.0, sha256
graph        feaa0155b64d78fe529d805c6e17430fb3ce9fe1c5b5900eb6e267e2fa077294
Manual       commit used to build the graph
             09b20f6fd2714a44b974d4ba38612f4dfdc73c11
DualGauge    code commit
             833b3fd0167106fb581e084c1bd71be4e389e5d7
BaxBench     v1.0.0 (agent runs) a67abd9; main branch (direct call) de885cd9
\end{verbatim}

\subsection{Appendix F. The two
benchmarks}\label{appendix-f.-the-two-benchmarks}

\textbf{DualGauge} {[}1{]}. The benchmark has 307 tasks in Python, C++
and JavaScript; 177 are in Python. Each task gives a functional
specification as text, with functional tests and security tests. The
tasks are short: a function or a small program, and in our sample 31 of
the 59 tasks use a web framework. An executor runs the generated code in
a sandbox, and a language-model judge (\texttt{gpt-4o} by default)
scores each test result as PASS, FAIL or UNKNOWN; only PASS counts. The
benchmark reports three task-level metrics: pass@1 (all functional tests
pass), secure@1 (all security tests pass) and secure-pass@1 (both). We
used the paper version arXiv:2511.20709v2 and code commit
\texttt{833b3fd}. The authors report that direct generation by the
strongest model stays below 15\% joint success in every language, and
that the agentic systems they tested gave no advantage over direct
generation.

\textbf{BaxBench} {[}10{]}. The benchmark has 28 backend scenarios, each
in 14 environments (frameworks and languages), for 392 tasks. Each
scenario gives an OpenAPI specification of the application. The
benchmark runs functional tests and then exploits against the
application in containers; each security test checks for one or more
weaknesses, and the scenarios probe 13 weakness types in total. A task
is \emph{correct} if it passes the functional tests, and \emph{correct
and secure} if, in addition, no exploit succeeds. The authors publish
results for three prompts: No Security Reminder, Generic Security
Reminder (``follow security best practices'') and Oracle Security
Reminder. They describe the last one as an ``unrealistic oracle security
reminder, where the developer anticipates all the security
vulnerabilities associated with the scenario and gives specific
instructions to avoid them''. They add that obtaining this knowledge
``is highly non-trivial, and a priori often impossible in practice'',
and that they include it ``only to gain an understanding of the upper
bound on the achievable security performance solely through prompting''.
We used BaxBench v1.0.0 (\texttt{a67abd9}) in the agent runs and the
main branch (\texttt{de885cd9}) in the direct-call run. The published
values that we cite for Claude Opus 4.5 with extended thinking (mean
over the 14 environments; leaderboard files read on 2026-10-04) are:

\needspace{7\baselineskip}

\begin{longtable}[]{@{}lrr@{}}
\toprule\noalign{}
Prompt & Correct and secure & Correct \\
\midrule\noalign{}
\endhead
\bottomrule\noalign{}
\endlastfoot
No Security Reminder & 0.5612 & 0.8622 \\
Generic Security Reminder & 0.6607 & 0.8469 \\
Oracle Security Reminder & 0.6862 & 0.8087 \\
\end{longtable}

\subsection{Appendix G. The SbD-ToE manual, knowledge graph and MCP
server}\label{appendix-g.-the-sbd-toe-manual-knowledge-graph-and-mcp-server}

\textbf{Manual.} The SbD-ToE manual is a security-by-design manual in 14
chapters that cover the development life cycle {[}26{]}. It classifies
applications in three levels of criticality, L1 to L3, and it states
requirements, controls and policies for each level. Its requirements
chapter relates the requirements to frameworks such as the OWASP ASVS
and the NIST SSDF. The knowledge graph used here was built from the
manual at commit \texttt{09b20f6fd271...}, as recorded in the payload
that the server returned (Appendix A).

\textbf{Knowledge graph.} The graph holds the requirements with stable
identifiers, the levels at which they apply, the controls that implement
them, mechanisms, practices, artefact types and evidence patterns, and
the links between them. The server release 0.10.1 ships the graph
release v1.5.0 (\texttt{sha256\ feaa0155b64d...}, built on 2026-06-12).

\textbf{Server.} The server is the npm package
\texttt{@shiftleftpt/sbd-toe-mcp} 0.10.1
(\texttt{dist.shasum\ 3f775ef9de8f...}). Its tool
\texttt{prepare\_sbd\_toe\_codegen\_context} takes three inputs:
\texttt{task} (the task text), \texttt{mode} (here \texttt{codegen}) and
\texttt{risk\_level} (here \texttt{L2}). It returns the activation
trace, the activated scope (requirements, control objectives, controls,
mechanisms, practices, artefact types and evidence patterns), a citation
map of the identifiers, a completeness report and rules for using the
answer. Appendix A shows one answer, abridged.

\begin{center}\rule{0.5\linewidth}{0.5pt}\end{center}

\subsection{References}\label{references}

{[}1{]} R. Patir, K. Guo, S. Barua, A. Pathak, D. Gudimetla, J. Guo, H.
Hu, H. Cai. \emph{DualGauge: Automated Joint Security-Functionality
Benchmarking of Specification-Only Code Generation by LLMs and Coding
Agents}. arXiv:2511.20709v2, 2025.

{[}2{]} P. Farinha. \emph{AppSec Core v1: A Formalized Normalization
Ontology for Application Security}. OSF preprint, 2026.
doi:10.17605/OSF.IO/U9CRD.

{[}3{]} W3C OWL Working Group. \emph{OWL 2 Web Ontology Language
Document Overview (Second Edition)}. W3C Recommendation, 2012.

{[}4{]} H. Knublauch, D. Kontokostas (eds.). \emph{Shapes Constraint
Language (SHACL)}. W3C Recommendation, 2017.

{[}5{]} Model Context Protocol. \emph{Model Context Protocol
Specification}, version 2026-07-28.
\url{https://modelcontextprotocol.io/specification/2026-07-28} (accessed
2026-10-06).

{[}6{]} H. Pearce, B. Ahmad, B. Tan, B. Dolan-Gavitt, R. Karri.
\emph{Asleep at the Keyboard? Assessing the Security of GitHub Copilot's
Code Contributions}. arXiv:2108.09293, 2021 (IEEE S\&P 2022).

{[}7{]} N. Perry, M. Srivastava, D. Kumar, D. Boneh. \emph{Do Users
Write More Insecure Code with AI Assistants?} ACM CCS 2023.

{[}8{]} J. Peng, L. Cui, K. Huang, J. Yang, B. Ray. \emph{CWEval:
Outcome-driven Evaluation on Functionality and Security of LLM Code
Generation}. LLM4Code 2025. arXiv:2501.08200.

{[}9{]} J. Chen et al.~\emph{SecureVibeBench: Benchmarking Secure Vibe
Coding of AI Agents via Reconstructing Vulnerability-Introducing
Scenarios}. ACL 2026. arXiv:2509.22097.

{[}10{]} M. Vero, N. Mündler, V. Chibotaru, V. Raychev, M. Baader, N.
Jovanović, J. He, M. Vechev. \emph{BaxBench: Can LLMs Generate Correct
and Secure Backends?} ICML 2025 (Proceedings of the 42nd International
Conference on Machine Learning, PMLR). arXiv:2502.11844v3.

{[}11{]} J. He, M. Vechev. \emph{Large Language Models for Code:
Security Hardening and Adversarial Testing}. ACM CCS 2023.

{[}12{]} J. He, M. Vero, G. Krasnopolska, M. Vechev. \emph{Instruction
Tuning for Secure Code Generation}. arXiv:2402.09497, 2024.

{[}13{]} J. Shi, T. Zhang. \emph{RESCUE: Retrieval Augmented Secure Code
Generation}. ICLR 2026. arXiv:2510.18204.

{[}14{]} X. Xu, L. Wu, J. te Lintelo, S. Zhang, K. Gai, S. Picek.
\emph{Activating Latent Security Knowledge through LLM-Guided Risk
Analysis for Secure Code Generation}. arXiv:2606.16244v2, 2026.

{[}15{]} A. Nunez, N. T. Islam, S. K. Jha, P. Najafirad.
\emph{AutoSafeCoder: A Multi-Agent Framework for Securing LLM Code
Generation through Static Analysis and Fuzz Testing}. NeurIPS 2024
Workshop on Safe \& Trustworthy Agents. arXiv:2409.10737.

{[}16{]} R. N. H. Khan, D. Wasif, J.-H. Cho, A. Butt. \emph{Multi-Agent
Code-Orchestrated Generation for Reliable Infrastructure-as-Code}.
arXiv:2510.03902, 2025.

{[}17{]} ISO/IEC 25010:2023. \emph{Systems and software engineering ---
Systems and software Quality Requirements and Evaluation (SQuaRE) ---
Product quality model}. 2023.

{[}18{]} M. Souppaya, K. Scarfone, D. Dodson. \emph{Secure Software
Development Framework (SSDF) Version 1.1}. NIST SP 800-218, 2022.

{[}19{]} L. Zheng et al.~\emph{Judging LLM-as-a-Judge with MT-Bench and
Chatbot Arena}. NeurIPS 2023 Datasets and Benchmarks Track.

{[}20{]} P. Wang et al.~\emph{Large Language Models are not Fair
Evaluators}. arXiv:2305.17926, 2023.

{[}21{]} P. Farinha. \emph{A Security-by-Design Evaluation Framework for
Ontology-Grounded Code Generation: A Pre-Registered Confirmatory Study
(Stage 1)}. Registered report, accepted in principle at ICSME 2026.
Pre-registration: OSF doi:10.17605/OSF.IO/H5AJE. SSRN 7003678, 2026.
\url{https://papers.ssrn.com/sol3/papers.cfm?abstract_id=7003678}.

{[}22{]} P. Farinha. \emph{What Must Be Done Here? Using a Bounded
Application-Security Ontology at the Point of Change}. Semantic Shields
3 @ JOWO 2026, use-case track. OSF doi:10.17605/OSF.IO/CAJ3P.

{[}23{]} P. A. Grassi et al.~\emph{Digital Identity Guidelines:
Authentication and Lifecycle Management}. NIST SP 800-63B, Revision 3,
2017 (updated 2020).

{[}24{]} M. Jones. \emph{JSON Web Algorithms (JWA)}. RFC 7518, IETF,
2015.

{[}25{]} ISO/IEC 7812-1:2017. \emph{Identification cards ---
Identification of issuers --- Part 1: Numbering system}. 2017.

{[}26{]} P. Farinha. \emph{Security by Design -- Theory of Everything
(SbD-ToE) manual}, chapter 01, ``Application Criticality
Classification''.
\url{https://www.securitybydesign.dev/en/sbd-toe/sbd-manual/classificacao-aplicacoes/intro},
2026 (accessed 2026-10-06). The knowledge graph used here was built from
the manual at commit \texttt{09b20f6fd271...} (Appendix E).

{[}27{]} OWASP Foundation. \emph{Application Security Verification
Standard (ASVS) 4.0.3}. 2021.

{[}28{]} SbD-ToE manual, source repository.
\url{https://github.com/SbD-ToE/sbd-toe-manual}. Documentation under CC
BY-SA 4.0; code under Apache 2.0 (accessed 2026-10-06).

{[}29{]} SbD-ToE MCP server, \texttt{@shiftleftpt/sbd-toe-mcp}.
\url{https://github.com/SbD-ToE/sbd-toe-mcp} and
\url{https://www.npmjs.com/package/@shiftleftpt/sbd-toe-mcp}. Apache 2.0
(accessed 2026-10-06).

\end{document}